\documentclass[useAMS,usenatbib,usegraphics]{mn2e}

\usepackage{natbib}
\usepackage{graphicx}
\usepackage{latexsym,amssymb}
\usepackage{amsmath}
\usepackage{gensymb}
\usepackage[draft]{hyperref}
\usepackage{hyperref}
\usepackage{capt-of}
\usepackage{textpos}
\usepackage{journalnames}
\usepackage{url}
\usepackage{multirow}
\usepackage{color}
\usepackage[T1]{fontenc}
\usepackage{float}
\usepackage{rotating}
\usepackage{tikz}
\usepackage{tablefootnote}
\usepackage{longtable}
\usepackage{footnote}
\usepackage{framed}
\usepackage{pgfgantt,rotating}
\usepackage{subfloat}
\usepackage{multirow}
\usepackage{blindtext}
\usepackage{booktabs}
\usepackage{tabu}
\usepackage{threeparttablex}

\hypersetup{
    bookmarks=true,         % show bookmarks bar?
    unicode=false,          % non-Latin characters in Acrobat�s bookmarks
    pdftoolbar=true,        % show Acrobat�s toolbar?
    pdfmenubar=true,        % show Acrobat�s menu?
    pdffitwindow=false,     % window fit to page when opened
    pdfstartview={FitH},    % fits the width of the page to the window
    pdftitle={My title},    % title
    pdfauthor={Author},     % author
    pdfsubject={Subject},   % subject of the document
    pdfcreator={Creator},   % creator of the document
    pdfproducer={Producer}, % producer of the document
    pdfkeywords={keyword1} {key2} {key3}, % list of keywords
    pdfnewwindow=true,      % links in new window
    colorlinks=false,       % false: boxed links; true: colored links
    linkcolor=black,          % color of internal links (change box color with linkbordercolor)
    citecolor=blue,        % color of links to bibliography
    filecolor=black,      % color of file links
    urlcolor=black,           % color of external links
    linkbordercolor={1 0 0},
    citebordercolor={0 1 0}
}

\makeatletter
\Hy@AtBeginDocument{%
  \def\@pdfborder{0 0 1}% Overrides border definition set with colorlinks=true
  \def\@pdfborderstyle{/S/U/W 0}% Overrides border style set with colorlinks=true
}
\makeatother

\title[Testing the existence of optical linear polarization in young brown dwarfs]{Testing the existence of optical linear polarization in young brown dwarfs \thanks{Based on observations of the Calar Alto Observatory (Spain), using CAFOS at the 2.2 m telescope.}}
\author[E. Manjavacas et al.]
{\parbox{\textwidth}{E. Manjavacas$^{1,2,3}$	\thanks{E-mail:  manjavacas@iac.es; elenamanjavacas@email.arizona.edu},
P. A.~Miles-P\'aez$^{1, 2, 4}$, 
M.~R.~Zapatero-Osorio$^{5}$,
B.~Goldman$^{6, 7}$,
E.~Buenzli$^{8}$, 
T.~Henning$^{7}$,
E.~Pall\'e$^{1, 2}$,
M.~Fang$^{3}$ 
}\vspace{0.4cm}\\
\parbox{\textwidth}{
$^{1}$Instituto de Astrof\'isica de Canarias, C/ V\'ia L\'actea, s/n, E38205 - La Laguna (Tenerife), Spain.\\
$^{2}$Dpt. de Astrof\'isica, Univ. de La Laguna, Avda. Astrof\'isico Francisco S\'anchez s/n, 38206 La Laguna (Tenerife), Spain.\\
$^{3}$Department of Astronomy/Steward Observatory, The University of Arizona, 933 N. Cherry Avenue, Tucson, AZ, 85721, USA.\\
$^{4}$The University of Western Ontario, Department of Physics and Astronomy, 1151 Richmond Avenue, London, ON N6A 3K7, Canada. \\
$^{5}$Centro de Astrobiolog\'ia  (CSIC-INTA).  Carretera de Ajalvir  km 4, E-28850 Torrej\'on de Ardoz, Madrid, Spain.\\
$^{6}$Observatoire astronomique de Strasbourg, Universit\'e de Strasbourg, 11 rue de  l'Universit\'e, F-67000 Strasbourg, France.\\
$^{7}$Max-Planck-Institut f\"ur Astronomie. K\"onigstuhl, 17. D-69117. Heidelberg, Germany.\\
$^{8}$Institute for Astronomy, ETH Zurich, Wolfgang-Pauli-Strasse 27, 8093 Zurich, Switzerland.}}

\begin{document}

%\date{In prep. for MNRAS}
%\date{Accepted 1988 December 15. Received 1988 December 14; in original form 1988 October 11}

\pagerange{\pageref{firstpage}--\pageref{lastpage}} \pubyear{2002}

\def\LaTeX{L\kern-.36em\raise.3ex\hbox{a}\kern-.15em
    T\kern-.1667em\lower.7ex\hbox{E}\kern-.125emX}

\maketitle

\label{firstpage}

\begin{abstract}
	{Linear polarization  can be used as a probe of the existence of atmospheric condensates in ultracool dwarfs. Models predict that the observed linear polarization increases with the degree of oblateness, which is inversely proportional to the surface gravity. We aimed to test the existence of optical linear polarization in a sample of bright young brown dwarfs, with spectral types between M6 and L2, {observable from the Calar Alto Observatory}, and cataloged previously as  low gravity objects using spectroscopy. Linear polarimetric images were collected in $I$ and $R$-band using CAFOS at the 2.2~m telescope in Calar Alto Observatory (Spain). The flux ratio method was employed to determine the linear polarization degrees. With a confidence of {3$\sigma$}, our data indicate that all targets have a linear polarimetry degree in average  below 0.69\% in the $I$-band, and below 1.0\% in the $R$-band,  at the time they were observed. We detected significant (i.e. $P/\sigma\ge3$) linear polarization for the young M6 dwarf 2MASS J04221413$+$1530525 in the $R$-band, with a  degree of $p^{*}=0.81\pm0.17$\%.}

\end{abstract}


%============================= section 1 =============================

\section{Introduction}\label{introduction}

Brown dwarfs are substellar objects that do not have enough mass to maintain stable hydrogen fusion in their cores, therefore, brown dwarfs cool with time, contracting and changing spectral types, from the M to Y spectral class \citep{Kirkpatrick2012}. Due to their evolution, the age and the mass are not determined by the spectral type, in contrast to stars. Thus, in addition to a spectrum, an accurate determination of brown dwarf ages is necessary to determine stellar masses \citep{Burrows1997}. 

Young brown dwarfs, with ages $\leq$500~Myr, are still contracting  to their final radii. Therefore, young brown dwarfs have lower gravity than mature field brown dwarfs ($>$500~Myr). Low gravity affects the chemistry  of the atmosphere of brown dwarfs, leading to the modification of some of the spectral features present in more evolved brown dwarfs: they have weaker alkali lines, and {in some cases redder colors and a triangular $H$-band \citep{Lucas, 2003ApJ...593.1074G}. {There are also very red  low mass dwarfs that do not show any evidence of youth or low gravity atmospheres \citep{Marocco_2014, Liu_Allers2016}. It has been suggested the the infrared flux excesses observed in both low- and high-gravity very red cool dwarfs are caused by extraordinary dusty atmospheres (\citealt{Looper_2008,  Barman_2011,  Gizis_2012, Allers_2016}), although other scenarios might be at play (e.g., different metallicity, presence of disks).     }

%\citet{Sengupta_Marley2010} (and references therein) predicted  that linear polarization in brown dwarfs caused by single and multiple scattering on dust particles in the range of 0.2--1\% in the optical and the near infrared. Nevertheless, if there is  horizontal asymmetry, the polarization originated from different areas of the surface will not cancel out, and linear polarization is expected. There are different sources of polarization in brown dwarf atmospheres: oblate atmosphere and  heterogeneous distribution of dust particles (even without oblate atmosphere).   lower gravities, rotationally induced non-sphericity is favored, therefore, for a similar amount of particles, we expect that the degree of polarization increases inversely with gravity. 
{Linear polarization is likely produced by scattering processes of the dusty particles  in brown dwarfs atmospheres \citep{Sengupta_Krishan2001, Sengupta2003, Sengupta2005}. Polarization can thus become a useful method to determine the properties of the atmospheric dust as the degree of polarization correlates with the particle size \citep{Sengupta_Krishan2001}.}
\citet[and references therein]{Sengupta_Marley2010}, predicted optical and near infrared values of linear polarization in the range 0-1\% for dusty ultracool dwarfs. Non zero linear polarization due to scattering processes is thought to arise as a result of the non cancellation of the polarization signal from different areas of the ultracool dwarf due to asymmestries such as an oblate shape or an heterogeneous distribution of dust.
At lower gravities, rotationally induced non-sphericity is favored. Therefore, for ultracool dwarfs with similar amounts of condensates, we expect that {the degree of linear polarization increases with the degree of oblatness}. 
Furthermore, {\cite{Marocco_2014}} and  \cite{Hiranaka2016} concluded that under the scenario of a \textit{dust haze} of particles in the upper atmospheres, the grains must have sub-micron sizes (typically between 0.15 and 0.4 $\mu$m) in order to explain the observed red colors.

Observationally, linear polarization in the optical and in the near infrared has been detected for some mature ultracool dwarfs, and also for young brown dwarfs \citep{Menard2002, Zapatero_Osorio2005, Goldman2009, Tata2009, Zapatero_Osorio2011, Miles_Paez2013, Miles_Paez2015}. The degree of polarization measured in field brown dwarfs in the optical is up to $\sim$1.4\%, which agrees with the theoretical predictions. 

In this paper,  we present the linear polarimetric photometry of a sample of well-selected low-gravity dwarfs whose masses are quite likely substellar.  By comparing our measurements with those of high-gravity dwarfs of related spectral types available in the literature, we will test whether there is a relation between linear polarization intensity and surface gravity. 

\section{Sample selection}\label{Sample_selection}

We selected  ultracool dwarfs { from \citet{Allers_Liu2013}}, with spectral types in the M--L transition (M6--L2), and confirmed as young objects, i.e. as low gravity brown dwarfs. All targets were observable from the Calar Alto observatory (Spain), and were bright enough  ($J\le$13.5~mag) in the $R$ and $I$-bands to achieve a precision of $\sigma_{P} \pm 0.2 \%$ with a 2.2~m telescope. Only six objects of \citet{Allers_Liu2013} sample satisfied our selection criteria. In Table \ref{literature} { we provide a list of the targets} with their magnitudes in $J$ and $W2$-bands, their spectral types, distances and their gravity flags. {Henceforth} we will use abridged names.

Our targets are brown dwarfs with low gravity signatures in their spectra: {weak alkali metal lines}, triangular $H$-band and redder colors than field brown dwarfs with the same spectral type. These objects have estimated ages lower than $\sim$500~Myr.

{2M J0045+1634 is classified as an L0$\beta$ by \citet{Faherty2016}, has lithium in its atmosphere and is a likely member of the Argus moving group according to Zapatero Osorio et al. (2014). Its age is estimated at the interval 10--100 Myr. \citet{Allers_Liu2013} classified this object as a very low gravity dwarf using near-infrared spectra, which is consistent with a young age. Both \citet{Miles_Paez2013} and \citet{Zapatero_Osorio2005} measured very low linear polarization indices, compatible with null polarization and in any case below 0.12\%, in the $J$ and $I$ bands.}

%Target 2M~J0045+1634 was classified by \cite{Faherty_2016} as a L0$\beta$, and was found to be a likely member of the Argus moving group ($\sim$ 40~Myr). However,  \cite{Allers_Liu2013} classified the object as a very low gravity dwarf using its near infrared spectrum. \cite{Zapatero_Osorio2005} calculated its degree of polarization in $I$ filter, obtaining $p^{*}$=0.00$\pm$0.12\%, and \cite{Miles_Paez2013} measured its polarization in the $J$ band to be $p^{*}$=0.00$\pm$0.11\%. 2M0045 is a likely member of the Argus moving group \cite{Zapatero_Osorio2014}, and has an age in the interval of 10--100~Myr, compatible with the presence of lithium in its cool atmosphere.

Object 2M~J0335+2342 is a M8.5 dwarf with Li detection at 6708~\AA   \citep{Reid2002}, which provides an upper limit on its age of $\sim$150~Myr for M dwarfs. It has intense and resolved H-$\alpha$ emission as well \citep{Shkolnik2009}. \cite{Allers_Liu2013} re-classified it as a young M7. \cite{Gagne2014} discussed that it might be a likely member of the $\beta$-Pictoris moving group.

The object 2M~J0422+1530 was classified as a M6$\gamma$ by \cite{Cruz} using its optical spectra. \cite{Allers_Liu2013} found that its near infrared spectrum was excessively red for its spectral type,{ and showed weak alkali metal lines}, indicating low surface gravity. {\citet{Faherty2012} reported an absolute parallax of 24.8$\pm$3.1~mas (d~=~40.3$\pm$5.0~pc), which is 6.4$\sigma$ discrepant with the absolute parallax reported by \citet{Liu_Allers2016} ($\pi_{abs}$~=~3.9$\pm$1~mas or d~=$\mathrm{240^{+70}_{-40}}$~pc).  2M0422 is  $\sim$10$^{\circ}$  to the South of the Taurus star-forming region on the sky (age $\sim$1~Myr, \citealt{Briceno1999, Luhman2004}), therefore, depending on which of the two values of the trigonometric distance reported is correct, 2M0422 might be before, embedded or behind the Taurus-Auriga star-forming region.  }

2M~J0443+0002 is a L0 dwarf classified as a very low gravity object by \cite{Allers_Liu2013}, showing all typical spectral characteristics of a low gravity dwarf. 

2M~J0602+3910 is a L2 intermediate gravity object \citep{Allers_Liu2013}, and a candidate member to the Pleiades moving group, with an estimated age of $\sim$100~Myr  \citep{Seifahrt2010}. 

{Finally, the object 2M J2057-0252 is a L2 intermediate gravity brown dwarf based on its low resolution spectrum \citep{Allers_Liu2013}. Its optical spectrum shows both lithium absortion and H$\alpha$ emission \citep{Cruz2003}.   \cite{Zapatero_Osorio2005} measured its linear polarization in $I$ band {obtaining a debiased value of} p*=0.00$\pm$0.38\%, and \cite{Miles_Paez2013} obtained a polarization in the $J$ band of p*=0.13$\pm$0.15\%.}

\begin{table*}
	\caption{List of observed targets with their magnitudes, spectral types and gravity characteristics.}  
	\label{literature}
	\centering
	\renewcommand{\footnoterule}{}  % to avoid a line before footnotes
	\begin{center}
		\begin{tabular}{lllllll}
			\hline 
			
			Name & $J$ [mag]  & $W2$ [mag] &   $d$ [pc] & SpT  & Gravity $\mathrm{flag^{a}}$&  Reference\\		
			
			\hline              
			2MASS J00452143+1634446 &   13.06$\pm$0.02& 10.39$\pm$0.02  &  17.5$\pm$0.6  & L0$\beta$      & VL-G, Li detection, 10-100~Myr {[9]}  &  1, 9 \\
			2MASS J03350208+2342356 &  12.25$\pm$0.02& 10.77$\pm$0.02   &  42.4$\pm$2.3   & M7               & VL-G, 10 Myr, Li detection {[7]}  & 2, 7 \\
			2MASS J04221413+1530525 &  12.76$\pm$0.02 & 10.71$\pm$0.02   & 40.3$\pm$5.0    & M6$\gamma$& VL-G & 3, 8 \\
													 &  						 & 								   &{ $240^{+70}_{-40}$ }   & {M6$\gamma$}& {VL-G} & {3, 11} \\
			2MASS J04433761+0002051 &  12.51$\pm$0.03 &  10.48$\pm$0.02  &     & L0                  & VL-G & 4 \\
			2MASS J06023045+3910592  & 12.30$\pm$0.02 &  10.13$\pm$0.02  &     & L2                  & INT-G, $\sim$100~Myr {[10]} & 5 \\
			2MASS J20575409--0252302 &  13.12$\pm$0.02 &  11.02$\pm$0.02  &  14.2$\pm$0.8   & L2                  & INT-G  & 6, 8  \\
			
			\hline

		\end{tabular}
	\end{center}
	
	\begin{tablenotes}
		\small
		\item References: [1] - \cite{Kendall2004}, [2] - \cite{Gizis2000}, [3] - \cite{Reid1995}, [4] - \cite{Hawley2002}, [5] - \cite{Lepine2002}, [6] - \cite{Menard2002}, [7] - \cite{Shkolnik2012}, [8] - \cite{Faherty2012}, [9] - \cite{Zapatero_Osorio2014}, [10] - \cite{Allers_Liu2013}, [11] - {\cite{Liu_Allers2016}} \\
		
		\item a: {Gravity class classification provided by \cite{Allers_Liu2013}: VL-G: the object has spectral characteristics consistent with very low gravity; INT-G: the object has spectral characteristics consistent with  intermediate gravity dwarfs.}
	\end{tablenotes}

\end{table*}

%------------------------------------------

% Add table with targets and data from the literature (like in Bonnefoy et al. 2013)

%------------------------------------------

\section{Observations and data reduction}\label{data_reduction}

We collected linear polarimetric images using the Calar Alto Faint Object Spectrograph (CAFOS, \citealt{Patat2011}), at the 2.2~m telescope at the Calar Alto Observatory (Spain). It is mounted at the Cassegrain focus, and it is equipped with a Wollaston prism for polarimetry that provides an effective beam separation of $\sim$20", plus a half-wave retarder plate. With this combination we were able to measure linear polarization  using dual-beam imaging polarimetry. CAFOS has a 2048 x 2048 pixel SITe CCD detector, with a scale of 0."53/pixel. The CCD was windowed to the central 1024 x 1024 pixels, with a field of view of 9'~x~9'. Observations were obtained during the whole nights between October the 23th and October the 27th  2014. 

Images were obtained in $R$ (641~nm, passband 158~nm) and $I$ bands (850~nm, passband 150~nm). All targets were observed in both filters in different days. Bias calibration frames, and twilight and dome flat frames with the polarization optics were taken every day in $I$ and $R$ filters. No systematic variation was detected in the calibration frames from night to night. Beside our targets, we observed two types of standard stars at approximately the same CCD spot (512, 512 on the windowed images): {two polarized {standard stars}, BD+25~727 and HD251204, and  two non-polarized {standard stars}, G191B2B and BD+28~4211. To investigate  instrumental polarization we used only G191B2B, because the other unpolarized star has a close visual companion that may affect the  polarization measurements . The white dwarf G191B2B allowed }us to test the stability of the instrument and the existence of instrumental polarization, {while the polarized standard stars allow us to identify zero offsets}.

We collected images at four different angles of the retarder plate: 0$^{\circ}$, 22.5$^{\circ}$, 45$^{\circ}$, and 67.5$^{\circ}$. The observing log is shown in Table  \ref{log0}, including: date of observation for each target, filter of observation, exposure time, number of exposures per each retarder plates position, airmass, aperture and FWHM (Full Width at High Maximum). % In Table \ref{log0} we provide the dates of the observations in the different filters and per position of the retarder plate, the exposure times for each position, and the airmass at the time of the observation. We observed several standard stars to perform calibrations on the measurement of polarization.

We reduced the raw images using the $IRAF$ (\textit{Image Reduction and Analysis Facility}) standard routines. Raw images were bias-substracted and flat-field corrected using the corresponding images to their respective filter. We used the dome flats acquired as a part of the daily standard calibrations taken in the telescope.

%All the measurements of polarization of these objects were compatible with zero polarization within the 1$\sigma$ error bars. 

\begin{table*}
%	\small
	\caption{Linear polarimetric results of the observed targets.}  
	\label{results_table}
	\centering
	\renewcommand{\footnoterule}{}  % to avoid a line before footnotes
	\begin{center}
		\begin{tabular}{crrrrrrrr}
			\hline
			
			Name & Filter & Obs. date & MJD & $q$ (\%)  & $u$ (\%) &  {$P$} (\%){$\mathrm{^{a}}$} & $p*$ (\%)$\mathrm{^{a}}$ & $\theta$ ($^{\circ}$) \\		
			
			\hline              
			2MASS J0045+1634 & I & 20141024 & 56954.9929000 & $-0.57\pm0.20$ &   $0.13\pm0.10$ &     $0.59\pm0.27$  &   $0.52\pm0.27$  &    \\
			& R & 20141026 & 56956.9751000  & $0.02\pm0.57$ &  $0.43\pm0.34$  &   $0.43\pm 0.68$ &   $0.00\pm0.68$  &\\    
			& R & 20141027 &  56957.9586000 & $0.05\pm0.36$  & $-0.77\pm0.17$   &  $0.77\pm0.43$ &    $0.64\pm0.43$& \\
			\hline
			2MASS J0335+2342 &  I & 20141024 & 56954.9067000 &  $-0.22\pm0.04$ &  $ -0.15\pm0.05$ &      $0.27\pm0.16 $ &    $0.21\pm0.16$  &  \\
			& I & 20141027 & 56958.0300000 & $0.06\pm0.09$ & $ -0.03\pm0.06$   &  $0.07\pm0.19$ &    $0.00\pm0.19$ &   \\
			&  R & 20141025 &  56956.0211000 &  $-0.07\pm0.09$ &   $0.45\pm0.11$ &      $0.46\pm0.21$ & $0.41\pm0.21$    & \\
			& R & 20141026 & 56957.0715000 & $0.03\pm0.49$  & $-0.09\pm0.43$   &  $0.09\pm0.67$ & $0.00\pm0.67$ & \\
			\hline
			2MASS J0422+1530 &    I & 20141024 &  56954.9680000 & $0.19\pm0.07$ &  $0.19\pm0.07$   &  $0.27\pm0.18 $ &  $0.20\pm0.18 $ &   \\
			& I & 20141027 &  56958.1188000 &  $0.12\pm0.07$ &  $0.35\pm0.12$ & $0.37\pm0.21$ &    $0.31\pm0.21$ & \\
			& R & 20141025 & 56956.0914000 & $-0.64\pm0.06$  & $-0.53\pm0.07$ & $0.83\pm0.18 $ &    $0.81\pm0.17$ & $109.9\pm6.2$ \\
			\hline							
			2MASS J0443+0002 & I &  20141024 &  56955.0793000  & $0.06\pm0.04$ &  $-0.23\pm0.09$ &     $0.24\pm0.16$   &  $0.18\pm0.16$ & \\
			& I & 20141027 & 56958.0945000 &  $-0.13\pm0.23$  & $-0.01\pm 0.14$ & $0.13\pm0.31$ & $0.00\pm0.31$ &  \\
			& R & 20141025 &  56956.1636000 & $-0.69\pm0.24$  & $-0.34\pm0.32$  & $0.77\pm0.43$ &  $0.64\pm0.43$& \\
			\hline								
			2MASS J0602+3910 & I & 20141024 & 56955.1420000  & $0.21\pm0.11$  & $-0.12\pm0.17$ & $0.25\pm0.25$  &  $0.00\pm0.25$ & \\
			& I & 20141027 &  56958.1746000 & $-0.02\pm0.17$ &  $-0.05\pm0.15$ & $0.06\pm0.27$   & $0.00\pm0.27$ & \\
			&   R & 20141026 & 56957.1811000 & $-0.01\pm0.24$  & $-0.03\pm0.28$ & $0.04\pm0.39$  & $0.00\pm0.40$  & \\
			\hline								
			2MASS J2057--0252 & I & 20141024 & 56954.8426000 & $0.14\pm0.15$ &  $0.01\pm  0.13$ & $0.14\pm0.25$    & $0.00\pm0.25$ & \\								
			& I & 20141027 & 56957.8639000  & $ -0.03\pm0.18$  & $0.25\pm0.22 $ & $0.25\pm0.32$  & $0.00\pm0.32$ & \\
			& R & 20141025 & 56955.8381000 & $0.00\pm0.20$    &         $0.00\pm0.20$ & $0.00\pm0.32$  & $0.00\pm 0.32$ & \\
			\hline								
			G191B2B & I & 20141027    &  56958.5602780  & $-0.11\pm0.12$   & $-0.03\pm0.12 $ &  $0.11\pm0.17$ & $0.00\pm0.17$ & \\
			&  R &  20141025 &  56956.7283220  & $-0.11\pm0.18$  & $-0.10\pm0.10$  & $0.10\pm0.21$ & $0.00\pm0.21$  & \\
			&  R  &  20141026 & 56957.7343980  & $-0.03\pm0.10$ &  $-0.19\pm0.12$ & $0.19\pm0.15$  & $0.12\pm0.15$ & \\
			\hline				
			%			BD+28 &    &   &   &   &  & &  & \\
			BD+25 727 &  I  & 20141024  & 56955.6904170 &  $1.41\pm0.22$ & $5.55\pm0.17$  &$5.72\pm0.31$  & $5.71\pm0.31$ & $37.9\pm1.6$  \\
			\hline
			
			HD251204 &  I  & 20141024  & 56955.7046870    & $2.61\pm0.09$   & $-2.98\pm0.09$ &  $3.96\pm0.19$ & $3.96\pm0.19$& $155.7\pm1.4$  \\
			& I  & 20141027 & 56958.7065050 & $2.88\pm0.13$ & $-3.08\pm0.08$ & $4.21\pm0.21$  & $4.21\pm0.21$ & $156.5\pm1.4$  \\
			& R & 20141026  & 56957.6505560 &  $3.38\pm0.09$   &  $-3.28\pm0.03$ & $4.70\pm0.18$  & $4.70\pm0.18$ & $157.9\pm1.1$\\

			\hline

		\end{tabular}
	\end{center}
	
		\begin{tablenotes}
			\small
			\item {a: $P$ and $p*$ are always greater or equal  0.}
		\end{tablenotes}

\end{table*}

\section{Polarimetric analysis}\label{analysis}

We calculated the degree of linear polarization in the $I$ and $R$ bands, and the polarization angles by means of the Stokes parameters $q$ and $u$. The equations to compute the Stokes parameters, the degree of linear polarization, and the {angle of vibration of the polarization}  are taken from \cite{Zapatero_Osorio2011} and \cite{Miles_Paez2013}:

\begin{equation}
R^{2}_{q} = \frac{o(0º)/e(0º)}{o(45º)/e(45º)}  
\label{Rq}
\end{equation}

\begin{equation}
R^{2}_{u} = \frac{o(22.5º)/22.5(0º)}{o(67.5º)/e(67.5º)}  
\label{Ru}
\end{equation}

\begin{equation}
q = \frac{R_{q}-1}{R_{q}+1} 
\label{q}
\end{equation}

\begin{equation}
u = \frac{R_{u}-1}{R_{u}+1} 
\label{u}
\end{equation}

\begin{equation}
P =  \sqrt{q^{2}+u^{2}} \quad  
\label{P}
\end{equation}

\begin{equation}
%\quad   
\theta = 0.5 \tan^{-1}(u/q)
\label{theta}
\end{equation}

where $o$ refers to the flux of the ordinary beam, and  $e$  refers to the flux of the extraordinary beam in the dual images of the single frames. $P$ and $\theta$ are the linear polarization, and the angle of vibration of the linear polarization.

Fluxes of the targets and reference stars have been derived by doing aperture photometry using different aperture radii and  different annuli for the sky at different distances from the centroid of the target. We calculated the flux for every target employing circular photometric apertures of different sizes, from 0.5--6~$\times$~FWHM  with steps of 0.1~$\times$~FWHM, and 6 sky rings, which were annuli with inner radii of 3.5 through 6~$\times$~FWHM (steps of 0.5~$\times$~FWHM) and widths of 1~$\times$~FWHM. These {fluxes} were used to compute the Stokes parameters, $q$ and $u$. We chose the range of apertures in which their estimated values remained nearly flat  (typically 2--5~$\times$~FWHM, depending on brightness and filter, see Table \ref{log0}). For every night and target we obtained figures similar to  Figure \ref{aperture_fig}.  {All the measurements are compatible with zero polarization within {3$\sigma$}, suggesting that the instrument is free of instrumental polarization. The error obtained in the  $I$-band for the unpolarized stars was 0.17\%, and for the $R$-band was 0.18\%}. Uncertainties $u$ and $q$ are estimated as the standard deviation of values of the Stokes parameters $u$ and $q$. Uncertainties in $P$ are the quadratic sum of the uncertainties in $q$ and $u$, and the uncertainty derived from non-polarized standard star (0.17\% for the $I$-band and 0.18\% for the $R$-band).   The error in the polarization vibration angle is:

\begin{figure}
	\centering
	\includegraphics[width=9.1cm,height=8cm]{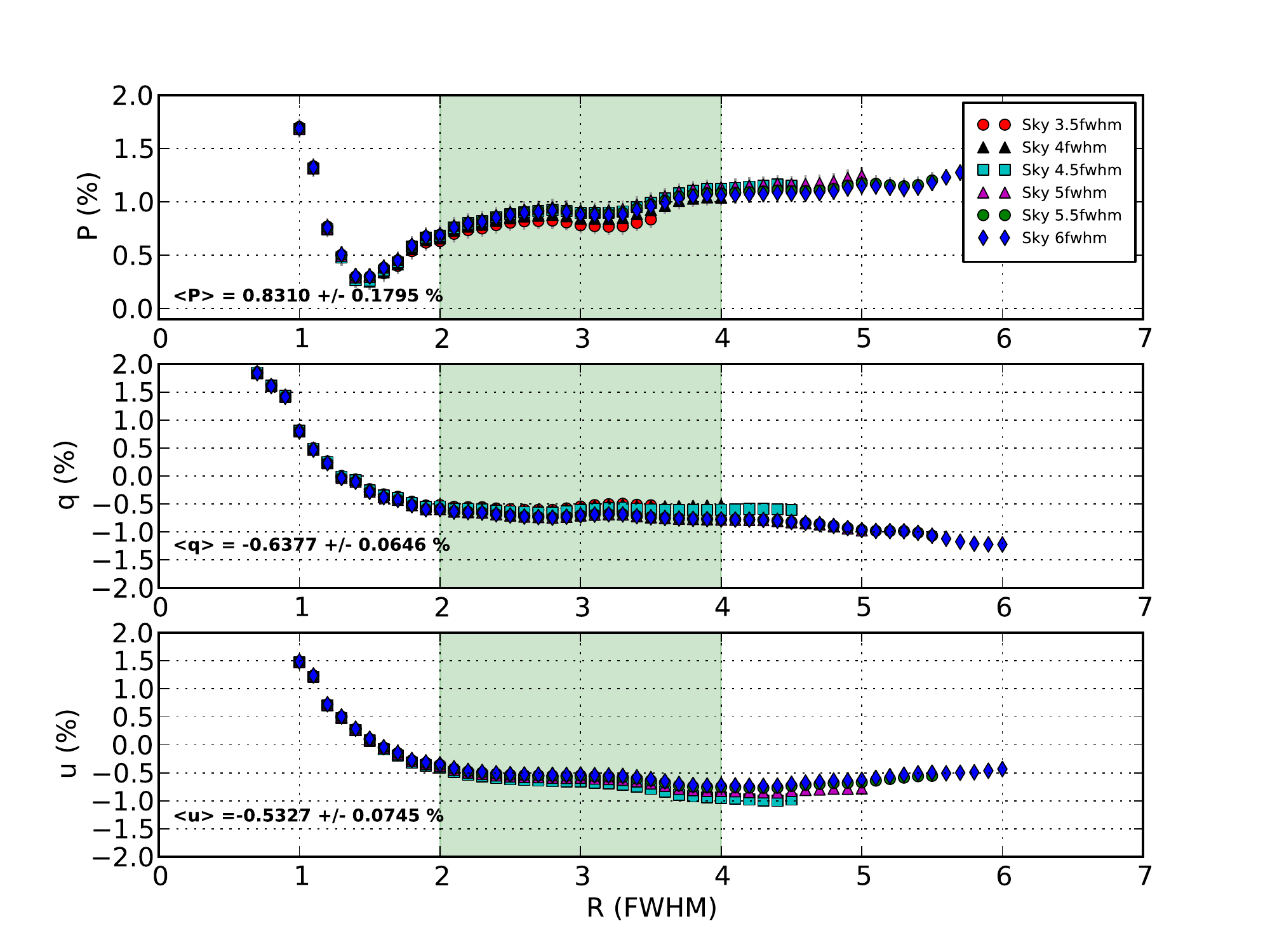}
	\caption{{Normalized Stokes parameters $q$ (middle) and $u$ (bottom), and the degree of linear polarization $P$ in {the $R$-band }as a function of the aperture radius (in FWHM unit) for the polarized object 2M0422. The green area denotes the region for which we averaged the Stokes parameters $q$ and $u$ (mean values also shown in each panel).  }}
	\label{aperture_fig}
\end{figure}

\small
\begin{equation}
\sigma_{\theta} = 28.65 \sigma_{P}/P
\label{Rq}
\end{equation}
\normalsize

Where $\sigma_{\theta}$  are the errors of the linear polarization angle in degrees, $\sigma_{P}$ is the uncertainty of the polarization degree. The factor 28.65\,$\sigma_{P}/P$ comes from the propagation of uncertainties, and it is only valid when $P/\sigma_{P}\geq$3 \citep{Serkowski1974, Wardle_Kronberg1974}.  {We identified as linearly polarized those sources in our sample that comply with  $P/\sigma_{P} \ge 3$ (i.e. 3$\sigma$ criterion)}.

{The degree of linear polarization is always a positive quantity, thus,} small values of $P$ and measurements affected by poor S/N are biased towards an overestimation of the true polarization. We applied \cite{Wardle_Kronberg1974} equation to debiase the linear polarization degree:

\small
\begin{equation}
p* = \sqrt{P^{2}-\sigma^{2}_{P}}
\label{Rq}
\end{equation}
\normalsize

We provide the observation dates, filter, Stokes parameters, linear polarization degree and debiased linear polarization degree with their respective uncertainties in Table~\ref{results_table}. Polarization vibration angles are only provided for those sources that satisfy the {3$\sigma$} criterion, and therefore are likely linearly polarized.

%  We highlighted in boldface the potentially polarized targets. 
%In our sample of six objects, only the target 2M~J0422+1530 in the $R$ filter satisfied the 3$\sigma$ criterion to be considered as a positive polarization detection. For the rest of the objects, the polarization obtained was compatible with zero linear polarization.

\section{Discussion}\label{Discussion}

%In our sample, we detected linear polarization only for target 2M0422 in the $R$-band, the rest of the targets have polarization compatible with zero (see Table \ref{results_table}).

All objects in our sample have a linear polarization degree in the $I$-band compatible with zero polarization. In the $R$-band, {only 2M0422 has a value of linear polarization} degree different from zero. The {3$\sigma$} upper limit on our detectability in the $I$-band is 0.69\%, and 1.0\%  in the $R$-band. {These upper limits were calculated by computing a mean of the uncertainties of all measurements of polarization for $I$ and $R$-bands respectively, and multiplying by 3 the obtained values}.  \cite{Sengupta_Marley2010} predicted a maximum polarization of 0.8\% in the $I$-band for a log\,g~=~4.5 object with $i$=90$^{\circ}$ and spectral type between M7 and L2. For a similar object but with $i$=30$^{\circ}$, the maximum predicted linear polarization is 0.2\%. Therefore, with the precision of our measurements, we would only be able to detect polarization for objects with $i$ close to 90$^{\circ}$.

 {The positive detection of polarization on the $R$-band and not in the $I$-band  might be due to the size of the particles producing the polarization, or to time variable polarization, depending on the mechanisms producing the polarization in 2M0422.}
 
 %{Regarding 2M0422, the positive detection of strong polarization in the $R$-band and no significant polarization in the $I$-band, despite the proximity of the two central wavelengths, implies that the polarization is maximum at a wavelength $\le$400 nm following the emprical function of the Serkowski's law of the interstellar medium polarization due to submicron particles \citep{Serkowski1975, Whittet1992}. We have adopted a 3$\sigma$ upper limit of $P(I) = 0.6\%$, which is three times the uncertainty associated with the polarimetric measurement of this particular source (Table~\ref{results_table}), and $K = 0.75$  after equation 3 of \citet{Whittet1992}. Alternatively, the quite different polarimetric measurements could be ascribed to time variable polarization intrinsic to 2M0422 (data were acquired on different nights). There is always the possibility of a rogue measurement. In what follows we assume that our measurements a correct within the quoted uncertainties.}
% The rotational period for brown dwarfs might be as long as 12~h \citep{Zapatero_Osorio2006}, but our observations in the different bands were $\sim$24~h apart (see Table \ref{log0} for further details)}.
{Below we describe} the most plausible mechanisms that can explain polarization found in 2M0422:

\begin{itemize}
	
	\item {Interstellar dust in the line of sight between the Earth and the targets. Interstellar dust might be a significant source of polarization for objects at distances further than {100}~pc {\citep{Tamburini2002, Bailey2010, Cotton2016}}.
		
		{If the distance derived by \citet{Faherty2012} is correct,  then 2M0422 cannot be embedded in the Taurus-Auriga star-forming region, thus, interstellar dust would not be a plausible explanation for the detected polarization for 2M0422.}
			
		{In the case in which the distance obtained by  \citet{Liu_Allers2016} is correct, then 2M0422 would be embedded or behind the Taurus-Auriga star-forming region. Therefore, the reddening found for 2M0422 colors, and the detected polarization in the $R$-band, could be explained by the presence of interstellar dust. This kind of linear polarization has an exponential dependence with wavelength  \citep{Serkowski1975}}. 
		{The positive detection of strong polarization in the $R$-band and no significant polarization in the $I$-band, despite the proximity of the two central wavelengths, implies that the polarization is maximum at a wavelength $\le$400~nm, following the empirical function of the Serkowski's law of the interstellar medium polarization due to submicron particles \citep{Serkowski1975, Whittet1992}. We have adopted a 3$\sigma$ upper limit of $P(I) = 0.6\%$, which is three times the uncertainty associated with the polarimetric measurement of this particular source (Table~\ref{results_table}), and $K = 0.75$  after equation 3 of \citet{Whittet1992}. }}

	%	 \item Intense magnetic fields can generate linear polarization by Zeeman effect or synchrotron emission. Brown dwarf outer layers have very low electric conductivity, this layer prevents the generation of magnetic fields \cite{Mohanty2002}. Therefore, we do not expect intense magnetic fields to be the cause of linear polarization for the objects in our sample.

	\item {Protoplanetary disk or debris disk surronding our target. Protoplanetary disks are expected to have lifetimes of about 10~Myr \citep{Luhman_2012}.  Some of the objects in our sample have estimated ages around $\sim$10 Myr. Thus, the linear polarization  detected in some of these objects may be originated by  a protoplanetary disk. Debris disks persist up to several Myr, but up to date, none has been confirmed around brown dwarfs.}
	
	In additon, in the case in which polarization is produced by a disk, we should find an excess of flux in the whole  mid and/or far-infrared, depending on the type of disk we might find. In Figure \ref{J_W2_SpT}, we show the color $J-W2$ as a function of the near infrared spectral type of our targets, and we compared them to the colors of the objects in \cite{Dupuy_Liu2012}, and the spectrophotometric relationship derived in the same work (blue line). Objects 2M~J0422 and 2M~J0045 are outside the {1$\sigma$} dispersion of the objects (dashed blue line), indicating mid-infrared flux excess in $J-W2$ color.
	
	{To try to confirm the existence of a disk, we searched for infrared excess in the whole SED (Spectral Energy Distribution) of 2M0422 using data available in VizieR \citep{Ochsenbein_2000}. After reddened the model atmospheric emission to fit its broad-band photometry ($A_{V}$=3.1~mag), we fitted the SED  predicted by the BT-Settl models for a M6 object, to the flux of 2M~0422+1530 in J, H, K (2MASS) and WISE (Fig. \ref{BD_SED}). No excess in the mid-infrared was found up to 11.6~$\mu$m. } 
	
	%We cannot discard the existence of debris disks around our targets, as debris disks have particles larger than 1.2~$\mu$m, and they are expected to produce polarization in the mid-infrared.
	
	\item Presence of a transiting companion. During the transit, the symmetry of the dusty atmosphere of the young brown dwarf or giant exoplanet is broken, and  the linear polarimetric signal should be enhanced \citep{Sengupta2016}. The peak polarization is predicted to range between 0.1--0.3\% in the near infrared. {\citet{Sengupta2016} estimated that the peak in polarization for L dwarfs due to the transit of an Earth-size or larger exoplanets vary in the range between 0.2\%--1.0\%. Our data were not conceived as a continuous monitoring program. Therefore, this scenario, although possibly less plausible than disk extinction or dusty clouds, cannot be discarded for 2M0422.}
	
	\begin{figure}
	%	\centering
		\includegraphics[width=9cm]{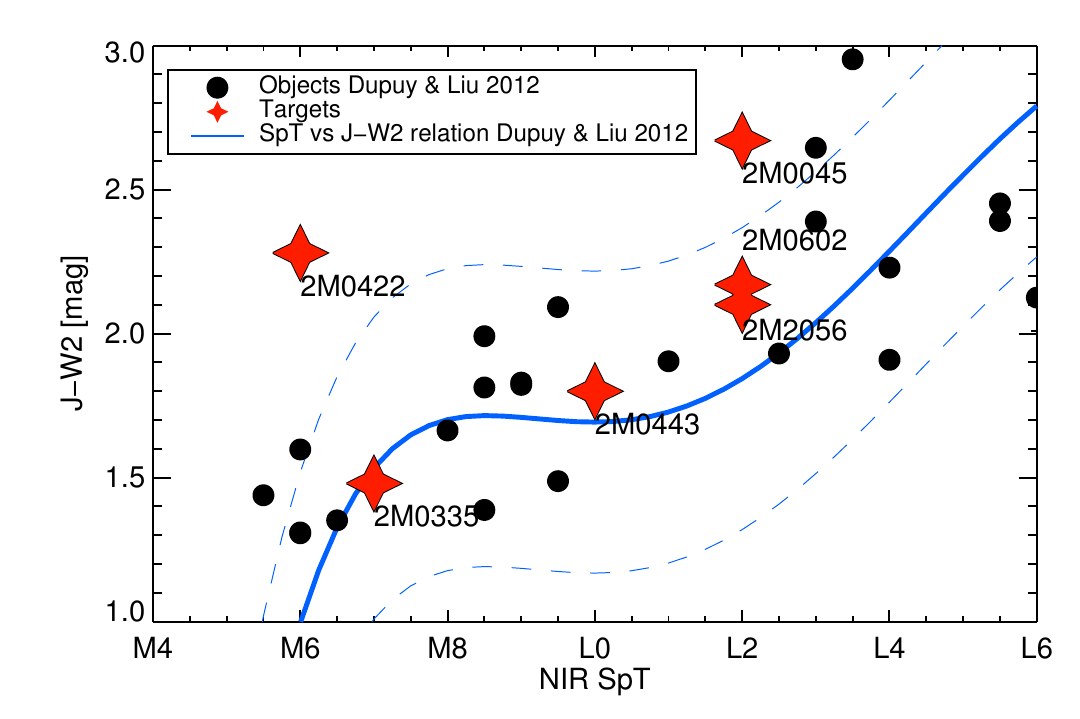}
		\caption{$J-W2$ color as a function of the near infrared spectral type. Our targets are plotted as red stars, objects from \citet{Dupuy_Liu2012} are plot as black dots, the spectrophotometric relationship derived by the same authors is plotted as a blue thick line, and the {1$\sigma$} limits are plot as a dashed thin blue line. {Uncertainties are smaller than the symbols}. Objects 2M0422 and 2M0045 are above the {1$\sigma$} limit, indicating mid-infrared excess, and  the possibility of the existence of a circumstellar disk.}
		\label{J_W2_SpT}
	\end{figure}

	\item Dust particles in the atmosphere of young brown dwarfs. L dwarf atmospheres are composed of clouds of iron and silicate grains of  $\mathrm{H_{2}O}$,  FeH and CO \citep{Cushing2005}, that produces absorption bands in the near infrared spectrum, and change the opacity of the atmosphere. The formation of the dust in ultracool dwarf atmospheres is influenced by surface gravity. 
	Low surface gravity is expected to enhance the formation of dust in the atmospheres of these objects, and therefore a higher degree of linear polarization is expected for young brown dwarfs. {As explained in Section \ref{introduction}, {\citet{Marocco_2014}} and \citet{Hiranaka2016} concluded that under the scenario of \textit{dust haze} of particles in the upper atmospheres, the grains must have sub-micron sizes (typically between 0.15 and 0.4 $\mu$m) in order to explain the observed red colors. Furthermore, that  the polarization index is greater in the R-band than in the I-band in the M6 J0422+1530 also  favors small characteristic particles, which better agrees with Hiranaka and {Marocco et al}.'s prediction. The detection of linear polarization in our targets would favor the scenario of the dusty atmospheres. {The quite different polarimetric measurements in $R$ and $I$-band could be ascribed to time variable polarization intrinsic to 2M0422, as the data were acquired on different nights. This is the most plausible scenario if the parallax reported by \citet{Faherty2012} is correct.}}
	
	%Low surface gravity is expected to enhance the formation of dust in the atmospheres of these objects, at there are more extended and the gas is cooler. 
	
	%	Furthermore, \citet{Sengupta_Marley2010} modeled dust clouds that may produce the observed polarization in L dwarfs, finding that the degree of polarization increases slowly with increasing oblatness, which varies inversely with gravity, as explained in Section \ref{introduction}. Therefore, we expect higher linear polarization degree for young brown dwarfs than for field brown dwarfs. 
\end{itemize}
%	Nonetheless, no polarization has been found in most of the targets of our sample, nor in $R$ neither $I$ band, with the exception of 2M~J0422 in the $R$-band. 

%	In {Figure \ref{SpT_polI}} we plot the degree of linear polarization in the $I$-band versus the near infrared spectral type. We plot the results for $I$-band polarization from \cite{Zapatero_Osorio2005} in black dots, plus the linear polarization results of \cite{Menard2002} (blue squares), together with our measurements (red stars).  Our targets follow the trend of the rest of the objects, although to confirm any tendency of the data, a higher statistic is needed.
% We observe that the later spectral type, the higher is the degree of the linear polarization in the $I$-band. 

\begin{figure}
	\centering
	\includegraphics[width=9cm]{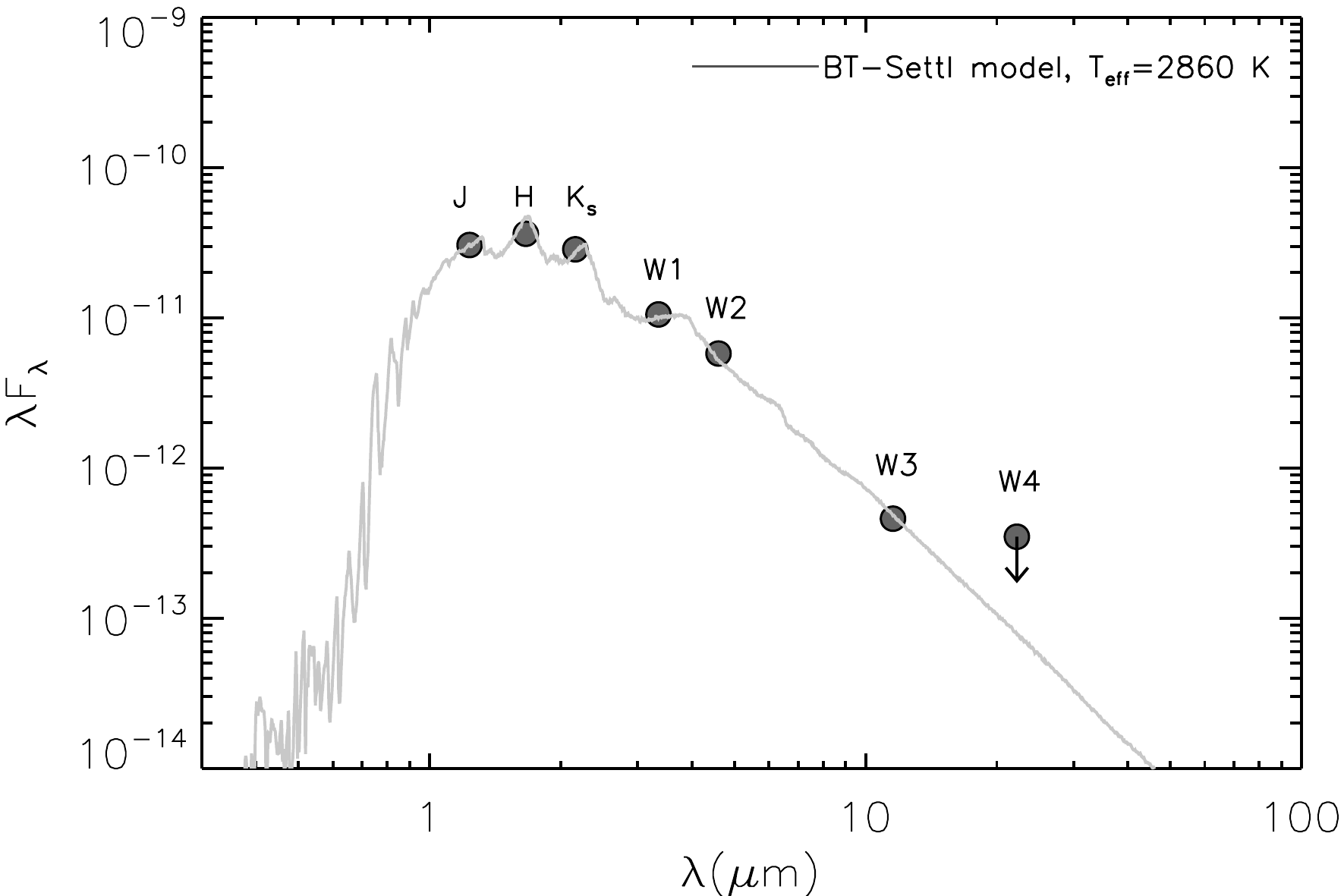}
	\caption{We fitted the 2MASS JHKs photometry of object 2M0422 to the  BT-settl atmospheric model with $\mathrm{T_{eff}}$=2860~K  (this is the $\mathrm{T_{eff}}$ corresponding to a M6 dwarf using the conversion from spectral type to $\mathrm{T_{eff}}$ from \citealt{Herczeg_Hillenbrand2014}). We employed two free parameters: the angular diameter $\theta$ of the source and the extinction ($A_{V}$) in the V band. We assumed a radius for 2M0422 of 1$R_{Jup}$. We used the extinction law of \citet{Cardelli1989} to redden model atmospheres, and adopted a total to selective extinction value typical of interstellar medium dust ($R_{V}=3.1$).  The fit give a visual extinction  for the object of $A_{v}\sim3.1$~mag. }
	\label{BD_SED}
\end{figure}

\section{Conclusions}

{In the present work, we aimed to test the existence of optical linear polarization in a sample of brown dwarfs that show signs of youth characteristics in optical or near infrared spectra. We selected the most extensive sample of  young brown dwarfs observable from CAHA, that were bright enough in the $R$ and $I$-bands to achieve a precision of $\sigma_{P} \pm 0.2 \%$ in polarimetry observation with a 2.2~m telescope}. We detected linear polarization for target 2MASS~J04221413+1530525 in the $R$-band ($p^{*}$=0.81$\pm$0.17\%). The two most plausible causes of the source of polarization in 2M0422 are:

\begin{itemize}
 
\item {Dust in the line of sight between the Earth and the object, only if we assume that the trigonometric parallax reported by \citet{Liu_Allers2016} is correct. In this case, 2M0422 might be embedded or behing the Taurus-Auriga star-forming region. This hypothesis would  explain the reddening of the spectrum of 2M0422, and the weak alkali lines indicating a young age of the object, in the case in which 2M0422 is a member of the star-forming region.} 

\item {A protoplanetary or debris disk, or dust particles in the atmosphere of the object, if we assume \citet{Faherty2012} as correct.  The last hypothesis would agree with {\citet{Marocco_2014}} and \citet{Hiranaka2016}   predictions about the sub-micron particles in young brown dwarf atmospheres, which would explain red colors and the detection of polarization in the $R$-band. }

\end{itemize}

For some of the objects in which polarization was not detected, the cause might be that a polarimetric cycle may cover a significant fraction of the rotation. If polarization is caused by inhomogeneous dusty concentrations over the surface, then it might be diluted over long exposures, {or it might be below our detection limits}.

%\citet{Tremblin2016} published a new version of atmospheric models for L dwarfs and extrasolar giant planets that proposed that no clouds are needed if the temperature gradient in the L dwarfs atmospheres is decreased, and non-equilibrium chemistry of $\mathrm{CO/CH_{4}}$ is taken into account. These models do not predict linear polarization for brown dwarfs. Nonetheless, further spectroscopic and photometric data are needed to further test this new version of atmospheric models.

To confirm the cause of the detected linear polarization in young brown dwarfs and in brown dwarfs in general, further polarimetric data with higher precision are necessary in the optical and in the near infrared. 
%Furthermore, {the current linear polarization models for brown dwarfs only consider the scenario of uniformly cloud coverage atmospheres. Nonetheless, several authors have probed through variability that some brown dwarfs may have patchy clouds or irregular cloud coverage \citep[and references therein]{Buenzli2014, Metchev2015, Ben2016}. In summary, further and higher precision linear polarimetric measurements, together with improved models are necessary to increase our statistics and to understand properly the origin of  the linear polarization detected in brown dwarfs. }

%\section{Conclusions}\label{Conclusions}

%--------------------------

	\section*{Acknowledgements}

{We are thankful to the referee P. Lucas, for his valuable comments.}

We gratefully acknowledge CAHA allocation time committee and CAHA Observatory staff for assiting the PI during the observations.  This work was supported by Sonderforschungsbereich SFB 881 "The Milky Way System" (subproject B6) of the German Research Foundation (DFG). This research has made use of the SIMBAD database, operated at CDS, Strasbourg (France), and IRAF, the \it{Image Reduction and Analysis Facility}.

%\newpage%%%%%%%%%%%%%%%%%%%%%%%%%%%%%%%%%%%%%%%%%%%%%%%%%%%%%%

%\begin{thebibliography}{}
%  \bibitem{Cushing2005} Author1, A.B., Author2, C.D.: 2001, AN 322, 1
%  \bibitem{} Author3, E.F., Author4, G.H.: 2001, AN 322, 10
% \bibitem{} Author5, I.: 2001, AN 322, 20
% \bibitem{} Author6, J.: 2001, AN 322, 30

\bibliographystyle{mn2e_fix}
\bibliography{BD_polarization}

%\end{thebibliography}

\appendix

\begin{table*}
	\footnotesize
	%	\begin{minipage}{18cm}
	\caption{Observing log. Exposures  format: exposure time (s) $\times$ number of circles in angle 0$^{\circ}$, exposure time (s) $\times$ number of circles in angle 22.5$^{\circ}$, exposure time (s) $\times$ number of circles in angle 45$^{\circ}$, exposure time (s) $\times$ number of circles in angle 67.5$^{\circ}$.}  
	\label{log0}
	\centering
	\begin{tabular}{lllllll}
		\hline
		
		Name & Date & Filter  & Exposures  & Airmass  & Aperture ($\times$~FWHM) & FWHM (") \\
		
		\hline              
		2M 0045+1634 & October 24, 2014 & I &  120$\times$4, 120$\times$4, 120$\times$4, 120$\times$4 & 1.15 & 2--5 & 2.3\\
		& October 26, 2014 & R &  300$\times$2, 300$\times$2, 300$\times$2, 300$\times$2   & 1.12 & 2--4 & 1.5 \\
		&								&    & 360$\times$1, 360$\times$1, 360$\times$1, 360$\times$1   &          &  & 1.7 \\
		& October 27, 2014 & I & 180$\times$3, 180$\times$3, 180$\times$3, 180$\times$3   & 1.06 &2--5 & 1.1  \\
		& October 27, 2014 & R & 300$\times$2, 300$\times$2, 300$\times$2, 300$\times$2  & 1.07 &2--4   & 1.1 \\
		&								&   & 420$\times$1, 420$\times$1, 420$\times$1, 420$\times$1   &        &    & 1.4  \\
		\hline	
		2M J0335+2342 & October 23, 2014 & R & 240$\times$1, 240$\times$1, 240$\times$1, 240$\times$1 & 1.04 &2--4  & 1.0 \\
		&							 &   & 300$\times$2, 300$\times$2, 300$\times$2, 300$\times$2 &			&  & 1.3  \\
		& October 23, 2014 & I & 120$\times$1, 120$\times$1, 120$\times$1, 120$\times$1  & 1.04 &2--5 & 1.4 \\
		&							 &   &  100$\times$2, 100$\times$2, 100$\times$2, 100$\times$2 &         & & 1.2 \\
		& October 24, 2014 & I &  100$\times$3, 100$\times$3, 100$\times$3, 100$\times$3 & 1.52 &2--5 & 1.3 \\
		& October 25, 2014 & R & 240$\times$1, 240$\times$1, 240$\times$1, 240  & 1.03 &2--4 &  1.3 \\
		&							 &     & 300$\times$1, 300$\times$1, 300$\times$1, 300$\times$1 &    & & 1.3 \\
		&							 &     & 260$\times$1, 260$\times$1, 260$\times$1, 260$\times$1 &    & & 1.3  \\
		& October 26, 2014 & R & 360$\times$3, 360$\times$3, 360$\times$3, 360$\times$3   & 1.05 &2--4 & 2.0  \\
		& October 27, 2014 & I &  150$\times$3, 150$\times$3, 150$\times$3, 150$\times$3  & 1.02 &2--5  &  1.4 \\
		
		\hline
		2M J0422+1530 & October 24, 2014 & I & 110$\times$2, 110$\times$2, 110$\times$2, 110$\times$2 & 1.40 &2--5  &2.9 \\
		& October 25, 2014 & R & 300$\times$2, 300$\times$2, 300$\times$2, 300$\times$2 & 1.07 &2--4 & 1.3\\
		&							 &    & 260$\times$1, 260$\times$1, 260$\times$1, 260$\times$1   &          & & 1.3 \\
		& October 27, 2014 & I & 180$\times$1, 180$\times$1, 180$\times$1, 180$\times$1    & 1.17 &2--5 & 2.3 \\
		&							 &   & 150$\times$3, 150$\times$3, 150$\times$3, 150$\times$3    &         & & 2.3 \\
		
		\hline
		2M J0443+0002 & October 24, 2014 & I & 150$\times$3, 150$\times$3, 150$\times$3, 150$\times$3   & 1.25 &2--5 & 3.4 \\
		& October 25, 2014 & R & 300$\times$2, 300$\times$2, 300$\times$2, 300$\times$2 & 1.40 &2--4 &  1.1\\
		&							 &     & 310$\times$1, 310$\times$1, 310$\times$1, 310$\times$1 &         &  & 1.2 \\
		& October 27, 2014 & I & 150$\times$1, 150$\times$1, 150$\times$1, 150$\times$1 & 1.25 &2--5 & 2.2 \\
		&							&     & 180$\times$1, 180$\times$1, 180$\times$1, 180$\times$1 &        &  & 2.3 \\
		
		\hline
		2M J0602+3910  & October 24, 2014 & I & 100$\times$4, 100$\times$4, 100$\times$4, 100$\times$4  & 1.00 &2--5 & 1.7 \\
		& October 26, 2014 & R  & 360$\times$3, 360$\times$3, 360$\times$3, 360$\times$3 & 1.02 &2--4 & 1.9 \\
		& October 27, 2014 & I  & 150$\times$3, 150$\times$3, 150$\times$3, 150$\times$3 & 1.10 &2--5 & 1.9 \\
		
		\hline
		2M J2057-0252 & October 24, 2014 & I & 120$\times$1, 120$\times$1, 120$\times$1, 120$\times$1  & 1.42 &2--5 & 3.9\\
		& 							&   & 150$\times$1, 150$\times$1, 150$\times$1, 150$\times$1   &		&  & 4.3 \\
		& 							&   & 180$\times$1, 180$\times$1, 180$\times$1, 180$\times$1   &		&  & 1.8 \\
		& October 25, 2014 & R & 300$\times$3, 300$\times$3, 300$\times$3, 300$\times$3  & 1.40 &2--4 & 1.7 \\
		& October 27, 2014 & I &  180$\times$2, 180$\times$2, 180$\times$2, 180$\times$2 & 1.50 &2--5 & 1.27 \\
		&								&   &  120$\times$1, 120$\times$1, 120$\times$1, 120$\times$1  &         &  & 1.2 \\

		\hline
		$\mathrm{G191B2B^{a}}$           & October 23, 2014 & I & 8$\times$1, 8$\times$1, 8$\times$1, 8$\times$1& 1.04 &2--6  & 1.0 \\
		& October 23, 2014 & R & 10$\times$1, 10$\times$1, 10$\times$1, 10$\times$1 & 1.04 &2--6 &0.9 \\
		& October 24, 2014 & I & 11$\times$3, 11$\times$3, 11$\times$3, 11$\times$3  & 1.12 &2--6 & 2.8\\
		& October 25, 2014 & R & 2.5$\times$2, 2.5$\times$2, 2.5$\times$2, 2.5$\times$2 & 1.60 &2--6 & 1.4\\
		& October 26, 2014 & R & 4$\times$3, 4$\times$3, 4$\times$3, 4$\times$3 & 1.19 &2--6 & 1.8 \\
		& October 27, 2014 & I & 4$\times$1, 4$\times$1, 4$\times$1, 4$\times$1  & 1.06 &2--6  & 1.4 \\
		&								&  & 4.5$\times$1, 4.5$\times$1, 4.5$\times$1, 4.5$\times$1 &    & & 1.4 \\
		&								&   & 5$\times$1, 5$\times$1, 5$\times$1, 5$\times$1 &  & & 1.4 \\
		& October 27, 2014 & R & 4$\times$3, 4$\times$3, 4$\times$3, 4$\times$3 & 1.06 &2--6 & 1.7 \\
		
		\hline
		BD+28 $\mathrm{4211^{a, c}}$       & October 25, 2014 & R & 1$\times$3, 1$\times$3, 1$\times$3, 1$\times$3 & 1.02 &2--6 & 1.2 \\
		& October 26, 2014 & R & 1.5$\times$3, 1.5$\times$3, 1.5$\times$3, 1.5$\times$3  & 1.28 &2--6 & 1.6 \\
		& October 27, 2014 & I & 2$\times$2, 2$\times$2, 2$\times$2, 2$\times$2 & 1.01 &2--6 & 1.1 \\
		&								&    & 2.5$\times$2, 2.5$\times$2, 2.5$\times$2, 2.5$\times$2 &    &  & 1.1 \\
		& October 27, 2014 & R &1.5$\times$3, 1.5$\times$3, 1.5$\times$3, 1.5$\times$3   & 1.01 &2--6 & 1.1 \\
		
		\hline
		BD+25 $\mathrm{727^{b}}$          & October 24, 2014 & I & 0.1$\times$3, 0.1$\times$3, 0.1$\times$3, 0.1$\times$3 & 1.12 &2--6  & 1.2 \\
		& October 27, 2014 & R & 0.5$\times$2, 0.5$\times$2, 0.5$\times$2, 0.5$\times$2 &  1.37 &2--6  & 2.9 \\
		& October 27, 2014 & I & 0.4$\times$3, 0.4$\times$3, 0.4$\times$3, 0.4$\times$3 &  1.32 &2--6 & 2.1 \\
		
		\hline
		$\mathrm{HD251204^{b}}$        & October 24, 2014 & I & 0.8$\times$2, 0.8$\times$2, 0.8$\times$2, 0.8$\times$2 &  1.05 &2--6 &1.1 \\
		& October 25, 2014 & R & 0.8$\times$3, 0.8$\times$3, 0.8$\times$3, 0.8$\times$3  & 1.07 &2--6 & 2.2\\
		& October 26, 2014 & R & 2$\times$3, 2$\times$3, 2$\times$3, 2$\times$3  & 1.03 &2--6 & 2.2 \\
		& October 27, 2014 & I & 0.9$\times$3, 0.9$\times$3, 0.9$\times$3, 0.9$\times$3  & 1.06 &2--6  & 2.2 \\
		
		\hline
		
	\end{tabular}
	
	\begin{tablenotes}
		\small
		\item Notes: a: Non-polarized standard star. b: Highly polarized standard star. {c: BD+28~4211 was not used as calibration star due to the presence of a nearby companion that may introduced spurious polarization.} 
	\end{tablenotes}
	
	%	\end{minipage}
	%	\tablefoot{\tablefoottext{a}{Non-polarized standard star}. \tablefoottext{b}{Highly polarized standard star}}
\end{table*}

%\section{This is the title of the first appendix}
%Larger tables, collections of images, spectra or similar kind of data shall be 
%presented in the appendix section rather than in the main body of the 
%text. Several appendices can be separated by the \verb+\section{+{\it title
%of appendix}\verb+}+ command. They are enclosed in the 
%\verb+appendix+ environment.

	\bsp

\label{lastpage}

\end{document}